\documentclass[osajnl,twocolumn,showpacs,superscriptaddress,10pt]{revtex4-1}
\usepackage[fleqn]{amsmath}
\usepackage{amssymb,graphicx}
\usepackage[english]{babel}
\makeatletter
\let\l@en\l@english
\makeatother

\begin{document}

\title{Nonlinear frequency conversion using high quality modes in GaAs nanobeam cavities}

\author{Sonia Buckley}\email{Corresponding author: bucklesm@stanford.edu} 
\author{Marina Radulaski}
\author{Jingyuan Linda Zhang}
\author{Jan Petykiewicz}
\affiliation{Ginzton Laboratory, Spilker Engineering and Applied Sciences Building, Stanford University, Stanford CA 94305}

\author{Klaus Biermann}
\affiliation{Paul-Drude-Institut f\"{u}r Festk\"{o}rperelektronik, Hausvogteiplatz 5-7 D-10117, Berlin, Germany}

\author{Jelena Vu\v{c}kovi\'{c}}
\affiliation{Ginzton Laboratory, Spilker Engineering and Applied Sciences Building, Stanford University, Stanford CA 94305}

\begin{abstract}
We demonstrate the design, fabrication and characterization of nanobeam photonic crystal cavities in (111)-GaAs with multiple high Q modes, with large frequency separations (up to 740 nm in experiment, i.e., a factor of 1.5 and up to an octave in theory).  Such structures are crucial for efficient implementation of nonlinear frequency conversion.  Here, we employ them to demonstrate sum frequency generation from 1300 nm and 1950 nm to 780 nm.  These wavelengths are particularly interesting for quantum frequency conversion between Si vacancy centers in diamond and the fiber optic network.\end{abstract}

\maketitle

\section{Introduction}

Photonic crystal cavities show great promise for nonlinear frequency conversion, due to their high quality (Q) factors and low mode volumes, which increase the conversion efficiency.  $\chi^{(3)}$ processes, such as Raman lasing \cite{takahashi_micrometre-scale_2013} and third harmonic generation \cite{Galli2010} in high Q silicon photonic crystal cavities, have been demonstrated, as well as the $\chi^{(2)}$ processes of second harmonic generation (SHG) and sum frequency generation (SFG) \cite{Mccutcheon2007,rivoire_second_2009, Galli2010, buckley_second_2013, Ota2013:Nanocavity-based, Buckley2014a}. Realization of low power, high efficiency frequency conversion devices could have important applications in quantum information processing. Target applications include on-chip frequency down-conversion of flying qubits, emitted by solid state emitters such as InAs quantum dots (QDs) or impurities in diamond, to telecommunications wavelengths \cite{mccutcheon_broadband_2009}, or up-conversion to the optimal frequency window for single photon detection \cite{langrock_highly_2005}. Additional applications include on-chip low power spectroscopic sources, up-conversion of mid-IR to visible for imaging applications \cite{zhou_ultrasensitive_2013}, and fundamental science studies such as strong coupling of single photons \cite{irvine_strong_2006} or single photon blockade \cite{majumdar_single-photon_2013}.  However, for $\chi^{(2)}$ three-wave mixing processes, which have relatively higher nonlinearities at lower powers, the ability to increase the efficiency at low power levels has been limited by the difficulty in engineering the large frequency separations between modes required for these processes.

One particularly promising candidate for achieving efficient nonlinear frequency conversion is  the 1D planar nanobeam photonic crystal cavity \cite{Deotare2009}, with Q factors of $>10^9$ simulated, and Q factors as high as $7.5\times10^5$ demonstrated \cite{Deotare2009}. Nanobeam designs for four wave mixing have been proposed \cite{lin_high-efficiency_2013} and demonstrated \cite{azzini_stimulated_2013}, and there have been additional proposals for terahertz frequency generation \cite{burgess_design_2009} and broadband frequency conversion of single photons \cite{mccutcheon_broadband_2009} using TE/TM polarized nanobeam cavities. Such TE/TM nanobeam cavities have been fabricated in silicon \cite{mccutcheon_high-q_2011}, but have yet to be implemented in high $\chi^{(2)}$ materials. Crossed nanobeam cavities \cite{rivoire_multiply_2011-1} offer another method for increasing the frequency separation of the modes involved in conversion, as the ability to individually select the wavelength of the two resonances is significantly improved and the thickness of the wafer is less critical; however such cavities suffer signficantly reduced Q factors as the frequency separation between modes is increased.

Here we demonstrate that the combination of lower and higher order TE modes in nanobeam cavities can be used for nonlinear frequency conversion.  We characterize nanobeam cavities fabricated in (111)B oriented GaAs membranes via fiber taper and cross polarized reflectivity. Finally we demonstrate sum frequency generation with two resonances separated by 700 nm.  We note that the use of (111) GaAs (as opposed to (001) GaAs) is necessary to enable conversion between three resonances (two for SHG), as discussed in our earlier work \cite{Buckley2014a}.

\section{Optimization of design}
\label{section:optimization}

\begin{figure*}
\includegraphics[width = 14cm]{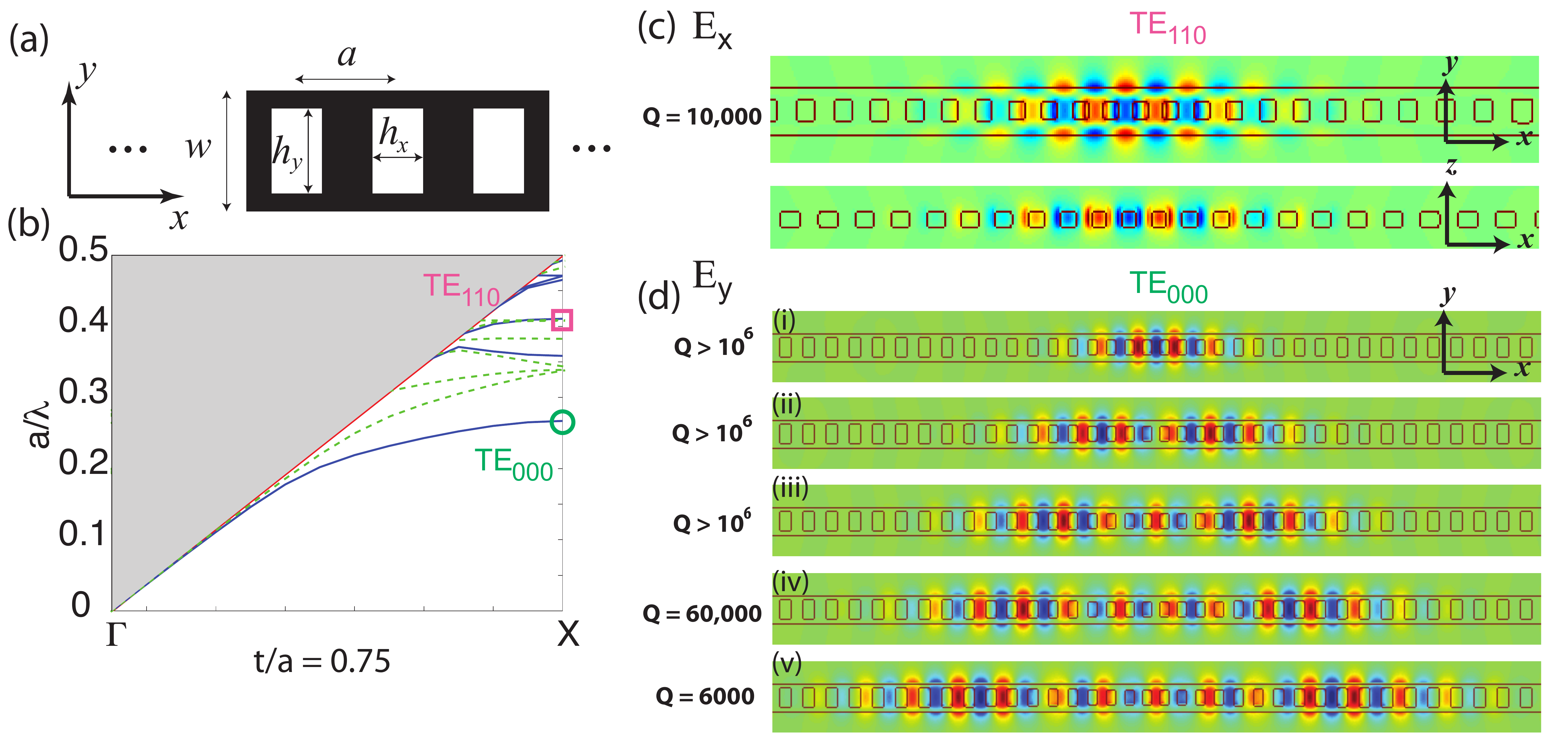}
\centering
{\caption{(a) Illustration of the periodic nanobeam.  (b) Band diagram for nanobeam with $h_y$ = 0.7 w, $h_x = 0.5a$, $n = 3.34$ for membrane thickness $t$ = 0.75a.  The blue solid lines show TE modes, the green dashed lines show TM modes.(c) The $E_x$ field component of the localized TE$_{110}$ mode used in this experiment. (d) The $E_y$ field component of the first to fifth order TE modes localized by the presence of the cavity from the TE$_{000}$ band in order of increasing frequency.}
\label{fig:simulation_fig}}
\end{figure*}

We first consider the band diagram of a nanobeam with lattice constant $a$, beam width $w =1.2a$, hole height $h_y = 0.7w$, hole width $h_x = 0.5a$, thickness $t/a$=0.75, and refractive index of 3.34 (GaAs at 1.8 $\mu$m) simulated with MIT photonic bands (MPB) \cite{johnson_block-iterative_2001} (Fig. \ref{fig:simulation_fig} (a) and (b)).  Transverse electric like (TE) modes are indicated by solid blue lines and transverse magnetic like (TM) modes by green dashed lines, and the light line indicated by the red line. A gentle perturbation in the periodic structure can localize one of these bands to the perturbed region.  For this experiment, we choose to confine modes from the TE$_{000}$ and TE$_{110}$ bands, as these have an appropriate symmetry and frequency separation.  The confinement is done using an adiabatic tapering of hole size and lattice constant.  The tapering is linear, with $a$, $h_x$ and $h_y$ tapering down to a minimum 340/430 of the original size at the center of the cavity.  The first five modes confined from the TE$_{000}$ band for a membrane thickness of $t/a = 0.75$, beam width $w/a = 1.2$, $h_y = 0.7 w$ and $h_x = 0.5 a$ are shown in Fig. \ref{fig:simulation_fig} (c). The modes alternate even and odd with respect to $x = 0$ and have decreasing Q factor and increasing extent (mode volume) with decreasing frequency. The mode volumes are 0.6$\left(\frac{\lambda}{n}\right)^3$, 1.1$\left(\frac{\lambda}{n}\right)^3$, 1.3$\left(\frac{\lambda}{n}\right)^3$, 1.6$\left(\frac{\lambda}{n}\right)^3$ and 1.9$\left(\frac{\lambda}{n}\right)^3$.  Similarly for the TE$_{110}$ band, several modes can be localized, and we obtain a Q factor of 10,000, which can be further optimized with the membrane thickness and hole size parameters \cite{buckley_inpreparation_2014}.  We note that this is not the only higher order mode of the nanobeam that is localized with high Q; the other modes in Fig. \ref{fig:simulation_fig} (b) can also be localized with high Q by optimization of various parameters, and these modes may also be used for other nonlinear frequency conversion processes, as will be discussed in more detail in ref. \cite{buckley_inpreparation_2014}.

\section{Linear characterization} \label{section:exp}
\begin{figure}
\includegraphics[width = 8.5cm]{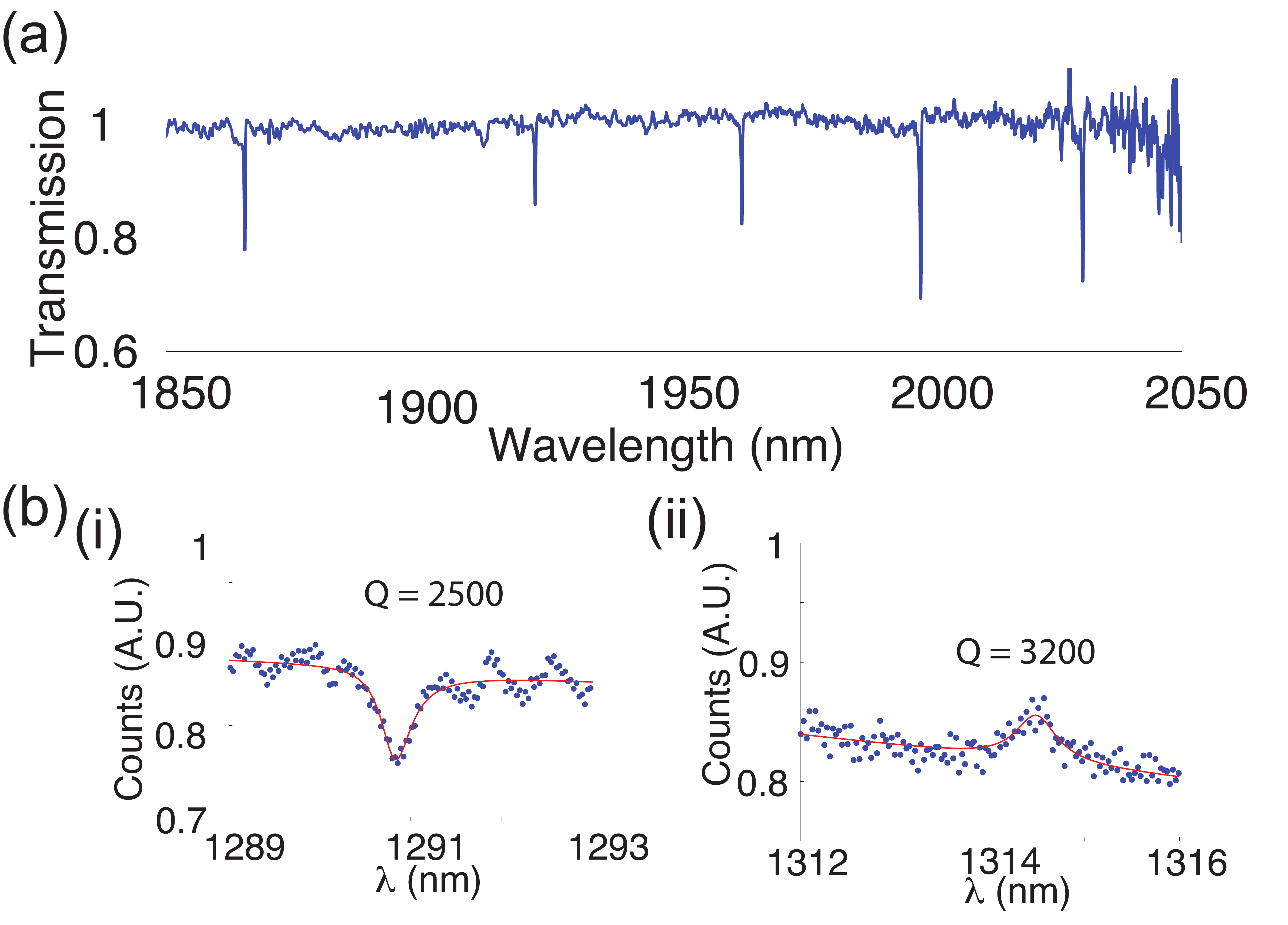}
\centering
{\caption{(a) Linear characterization of a nanobeam at longer wavelengths using the fiber taper method. (b) Cross polarized reflectivity measurement (free space) of the TE$_{110}$ modes of the nanobeam cavity in part (a).}
\label{fig:linear_mode_characterization}}
\end{figure}
We fabricate nanobeams in a 250 nm membrane in (111)B oriented GaAs with lattice constants $a$ from 450 nm to 650 nm (corresponding to $t/a$ = 0.56-0.38) and beam widths from 1.07 to 1.5 $a$ as described previously \cite{buckley_second_2013}.  The rest of the design follows section \ref{section:optimization}.  The highest frequency localized resonances of the TE$_{000}$ band of the nanobeam (which we will call the fundamental resonance) span from 1.45 $\mu$m to 1.87 $\mu$m with lattice constant variation, with the longest wavelength mode (the fifth confined mode of the band) measured at 2.03 $\mu$m.  Resonances up to 1.65 $\mu$m could be characterized via cross-polarized reflectivity  \cite{buckley_second_2013}. However, longer wavelength modes were characterized via fiber taper probing \cite{shambat_coupled_2010}, as the cross-polarized reflectivity method did not have high enough signal to noise.    In each case the cavity was probed with a broadband tungsten halogen white light source and reflected/transmitted signal detected on an extended InGaAs spectrometer.  An example of such a fiber taper measurement for a cavity, showing the first five confined TE$_{000}$ modes is shown in Fig. \ref{fig:linear_mode_characterization} (a) for a structure with lattice constant 620 nm ($t/a$ = 0.4).  The Q factors for the modes are all 8,000-10,000, and are limited by coupling to the fiber taper. 
We also characterized the higher order modes of the structure.  For the structure characterized in Fig. \ref{fig:linear_mode_characterization} (a), we observed two closely spaced higher order modes with Q of around 3000 via cross polarized reflectivity (Fig. \ref{fig:linear_mode_characterization} (b)).  Since the membrane thickness $t/a \approx 0.4$, we believe these modes to be the first two confined (even and odd) TE$_{110}$ modes of the structure.  The simulated even mode is shown in Fig \ref{fig:simulation_fig} (c), with a simulated Q factor of 10,000.  The highest Q higher order modes that we have measured have Q factors of around 8,000 at 1300 nm.  
\begin{figure}
\includegraphics[width = 8cm]{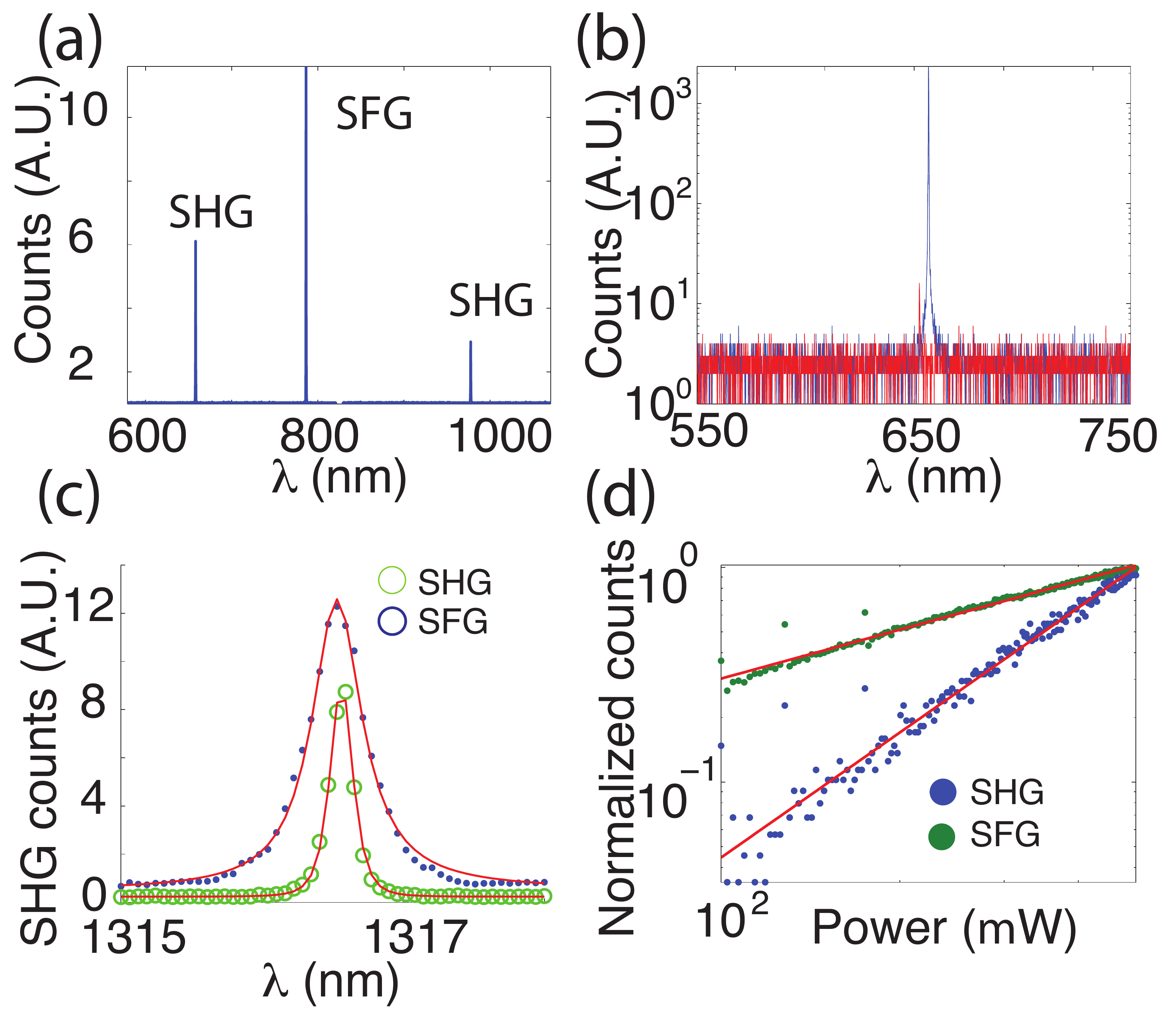}
\centering
{\caption{(a) SHG from the two modes (at 1314.5 nm and 1955 nm respectively), as well as SFG. (b) SHG is present even when the laser is not resonant with the cavity.  (c) Plot of second harmonic and sum frequency versus laser wavelength.  The fit is to a Lorentzian squared for SHG and Lorentzian for SFG. (d) As the power of the laser is scanned, the second harmonic follows a quadratic relation, while the SFG follows a sub-linear relation.}
\label{fig:Nonlinear_characterization}}
\end{figure}
\begin{figure}
\includegraphics[width = 8cm]{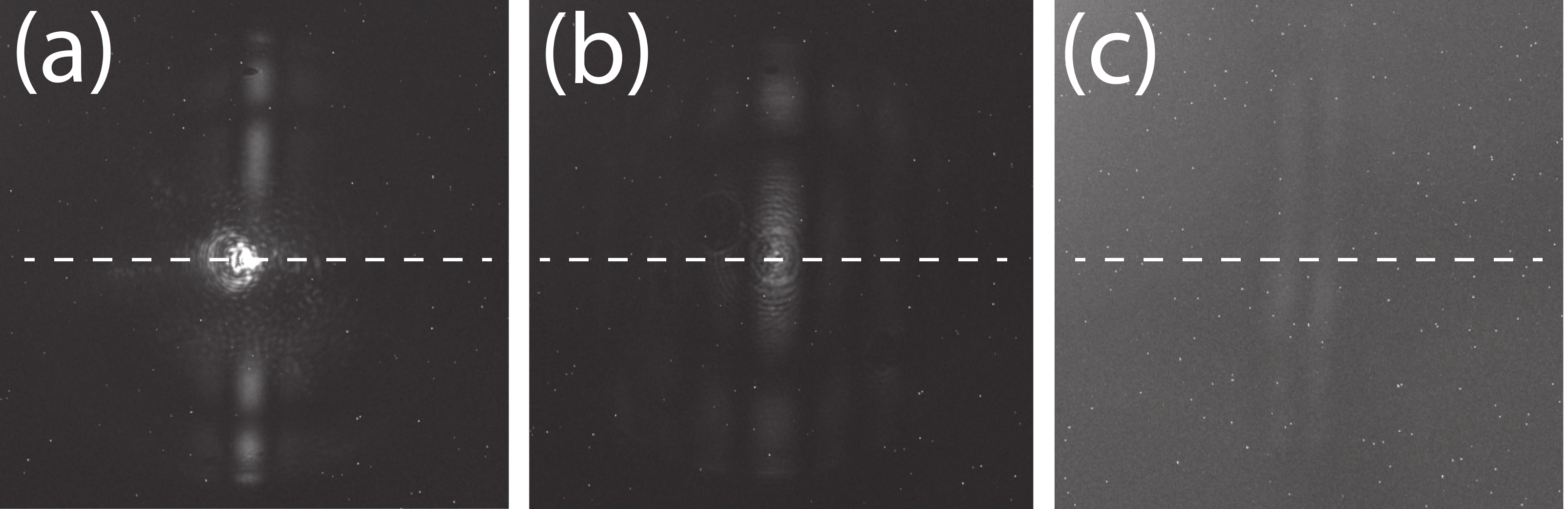}
\centering
{\caption{(a) k-space image of the second harmonic of the 1300 nm coupled laser.  (b) k-space image of the second harmonic from the 1955 nm laser.
(c) k-space image of the sum-frequency. The orientation of the nanobeam is indicated by the dashed white line.}
\label{fig:farfields}}
\end{figure}

\section{Sum frequency generation} \label{section:nonlinear}We next demonstrate SFG in a nanobeam photonic crystal cavity utilizing a confined TE$_{000}$ mode and a higher order mode of the nanobeam.  A tunable CW OPO is matched to the wavelength of the fourth confined TE$_{000}$ mode of the nanobeam shown in Fig. \ref{fig:linear_mode_characterization} (a) at 1955 nm, and is coupled to the mode at normal incidence through a high numerical aperture (NA = 0.95) objective lens (see ref. \cite{Buckley2014a} for details).  A Santec tunable diode laser is similarly matched to the higher order mode of the nanobeam at 1314.5 nm.  SHG and SFG from the two modes are collected back through the same objective lens when the laser is resonant with that mode. The SHG/SFG signal could either be sent to an imaging camera where either real space or momentum space was imaged, or to a spectrometer for spectral resolution. The second harmonic and sum frequency peaks are shown in Fig. \ref{fig:Nonlinear_characterization} (a). The tunable diode laser is scanned across the mode in order to verify that the mode is resonantly enhancing the second harmonic efficiency. Second harmonic from the nanobeam is observed even when the laser is not resonant with the cavity. However, there is a 200-fold enhancement in the signal when it is resonant with the mode of the cavity, as shown in Fig. \ref{fig:Nonlinear_characterization} (b).  This second harmonic is not seen generally; it can only be detected when the laser is on the nanobeam but not necessarily on the cavity (spatially or in frequency). This is perhaps due to enhanced in- and out- coupling, or surface SHG \cite{sanatinia_surface_2012, mccutcheon_resonant_2005} at the many additional etched surfaces of the GaAs.  A plot of the laser wavelength versus second harmonic counts and sum frequency counts is shown in Fig. \ref{fig:Nonlinear_characterization} (c).  The second harmonic counts are fit to a Lorentzian squared, while the sum frequency counts are fit to a Lorentzian, both indicating Q factors of between 3000-4000, which is also consistent with the cross-polarized reflectivity measurement on this structure. By scanning the power of the laser, we can plot the second harmonic counts and the sum frequency counts versus the laser power.  On a log-log plot, we obtain a straight line for the second harmonic counts, with a slope of 1.93, which is close to the expected value of 2. The sum frequency is expected to be a straight line with a slope of 1; however, we find that it is non-linear on both a log-log plot and a linear plot, and the best fit is for an exponent of 0.75.  This indicates nonlinear dependence on the laser power, which could be due to absorption or the laser power being diverted to SHG.  We also measure the experimental k-space distribution of the second harmonic from each of the cavity modes, in addition to the k-space distribution of the sum frequency.  This is shown in Fig. \ref{fig:farfields} with the position of the nanobeam indicated by the dashed white line.

\section{Discussion} The design presented in this paper could prove useful for frequency conversion between emitters (e.g. Si vacancy in diamond) and telecommunications wavelengths.  Our design can also be easily optimized for conversion between other wavelengths, such as 930 nm (InAs QD emission) and 1550 nm (optimal telecommunication transmission window). Other types of 1D or 2D waveguides could prove more suitable for engineering multiple high Q resonances with a large degree of overlap, where a gentle tapering or cavity confinement of these types of devices could prove fruitful \cite{lu_photonic_2013, logan_widely_2013}.  Future work could use inverse design \cite{lu_nanophotonic_2013} or genetic algorithms \cite{Minkov2014} to optimize resonances at widely spaced frequencies.  Even if the conversion efficiency per element is low, such doubly resonant microcavities could be linked together in a coupled cavity array to take advantage of both resonant cavity and slow light effects \cite{xu_propagation_2000}.  In order to achieve efficient frequency conversion, it is also important to engineer input and output coupling ports. For nanobeam photonic crystal cavities, this can be done via a side-coupled waveguide \cite{groblacher_highly_2013}.  Since the Q factor of the fundamental mode is so much higher than the Q of the second harmonic mode, the waveguide can be chosen to be critically coupled to the input mode without perturbing the second harmonic mode significantly.  Outcoupling of the second harmonic mode can be done via free space, or via a second waveguide with mirrors that prevent propagation at the fundamental wavelength but allow propagation at the second harmonic wavelength.\section{Conclusion}  We have demonstrated the design, fabrication and characterization of nanobeam photonic crystal cavities with multiple higher order modes with frequency separations of greater than 700 nm and spanning the range of 1300 nm to 2000 nm.  We use SFG with light input on these modes to upconvert light from 1300 nm to 780 nm by SFG with the signal at 1955 nm, and also demonstrate SHG from these long wavelength resonances to near infra-red wavelengths. Such devices are important for intracavity frequency conversion of light from integrated quantum emitters such as quantum dots and color centers in diamond. 

 \section*{Acknowledgements} Financial support provided by the Air Force Office of Scientific Research, MURI Center for multi-functional light-matter interfaces based on atoms and solids, National Science Graduate Fellowships, and Stanford Graduate Fellowships. This work was performed in part at the Stanford Nanofabrication Facility of NNIN supported by the National Science Foundation under Grant No. ECS-9731293, and at the Stanford Nano Center. J.V. thanks the Alexander von Humboldt Foundation for support.

\end{document}